# RADIATION-HARD OPTO-LINK FOR THE ATLAS PIXEL DETECTOR

K.K. GAN, K.E. ARMS, M. JOHNSON, H. KAGAN, R. KASS, C. RUSH, S. SMITH, R. TER-ANTONIAN, M.M. ZOELLER

*Department of Physics, The Ohio State University, Columbus, OH 43210, USA*

P. BUCHHOLZ, M. HOLDER, A. ROGGENBUCK, P. SCHADE, M. ZIOLKOWSKI

*Fachbereich Physik, Universitaet Siegen, 57068 Siegen, Germany*

The on-detector optical link of the ATLAS pixel detector contains radiation-hard receiver chips to decode bi-phase marked signals received on PIN arrays and data transmitter chips to drive VCSEL arrays. The components are mounted on hybrid boards (opto-boards). We present results from the opto-boards and from irradiation studies with 24 GeV protons up to 33 Mrad (1.2 x $10^{15}$ p/cm$^2$).

## 1. Introduction

The ATLAS pixel detector [1] consists of two barrel layers and two forward and backward disks which provide at least two space point measurements. The pixel sensors are read out by front-end electronics controlled by the Module Control Chip (MCC). The low voltage differential signal (LVDS) from the MCC is converted by the VCSEL (Vertical Cavity Surface Emitting Laser) Driver Chip (VDC) into a single-ended signal appropriate to drive a VCSEL. The optical signal from the VCSEL is transmitted to the Readout Device (ROD) via a fiber.

The 40 MHz beam crossing clock from the ROD, bi-phase mark (BPM) encoded with the data (command) signal to control the pixel detector, is transmitted via a fiber to a PIN diode. This BPM encoded signal is decoded using a Digital Opto-Receiver Integrated Circuit (DORIC). The clock and data signals recovered by the DORIC are in LVDS form for interfacing with the MCC.

The ATLAS pixel optical link contains 448 VDCs and 360 DORICs with each chip having four channels. The chips will be mounted on 180 chip carrier boards (opto-boards). Each opto-board contains seven optical links. The optical link will be exposed to a maximum total fluence of 2 x $10^{15}$ 1-MeV $n_{eq}$/cm$^2$ during ten years of operation at the LHC. We study the response of the optical link to 24 GeV/c protons where the expected equivalent dosage is 6.3 x $10^{14}$ p/cm$^2$. In this paper we describe the development of the opto-link, including the results from irradiations.

## 2. VDC Circuit

The VDC converts an LVDS input signal into a single-ended signal appropriate to drive a VCSEL in a common cathode array. The output current of the VDC is to be variable between 0 and 20 mA through an external control current, with a standing current (dim current) of ~1 mA to improve the switching speed of the VCSEL. The rise and fall times of the VCSEL driver current are required to be less than 1 ns. In order to minimize the power supply noise on the opto-board, the VDC should also have constant current consumption independent of whether the VCSEL is in the bright (on) or dim (off) state.

Figure 1 shows a block diagram of the VDC circuit. An LVDS receiver converts the differential input into a single-ended signal. The differential driver controls the current flow from the positive power supply into the anode of the VCSEL. The VDC circuit is therefore compatible with a common cathode VCSEL array. An externally controlled voltage, $V_{Iset}$, determines the current Iset that sets the amplitude of the VCSEL current (bright minus dim current), while an externally controlled voltage, tunepad, determines the dim current. The differential driver contains a dummy driver circuit which in the VCSEL dim state draws an identical amount of current from the positive power supply as is flowing through the VCSEL in the bright state. This enables the VDC to have constant current consumption.

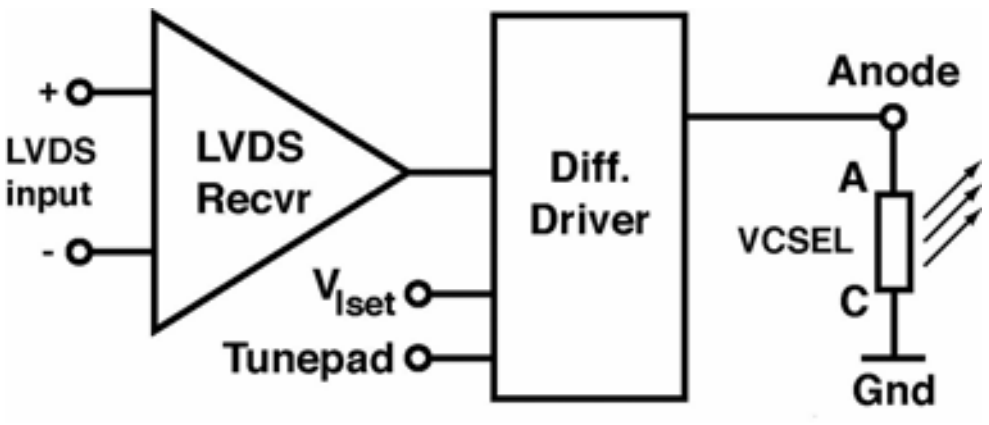

Figure 1: Block diagram of the VDC circuit.

## 3. DORIC Circuit

The DORIC decodes BPM encoded clock and data signals received by a PIN diode. The BPM signal is derived from the 40 MHz beam crossing clock by sending only transitions corresponding to clock leading edges. In the absence of data bits (logic level 1), this results simply in a 20 MHz signal. Any data bit in the data stream is encoded as an extra transition at the clock trailing edge.

The amplitude of the current from the PIN diode is expected to be in the range of 40 to 1000 µA. The 40 MHz clock recovered by the DORIC is required to have a duty cycle of (50 ± 4)% with a total timing error of less than 1 ns. The bit error rate of the DORIC is required to be less than $10^{-11}$ at end of life.

Figure 2 shows a block diagram of the DORIC circuit. In order to keep the PIN bias voltage (up to 10 V) off the DORIC, we employ a single-ended preamp circuit to amplify the current produced by the PIN diode. Since single-ended preamp circuits are sensitive to power supply noise, we utilize two identical preamp channels: a signal channel and a noise cancellation channel. The signal channel receives and amplifies the input signal from the anode of the PIN diode, plus any noise picked up by the circuit. The noise cancellation channel amplifies noise similar to that picked up by the signal channel. This noise is then subtracted from the signal channel in the differential gain stage. To optimise the noise subtraction, the input load of the noise cancellation channel should be matched to the input load of the signal channel (PIN capacitance) via an external dummy capacitance.

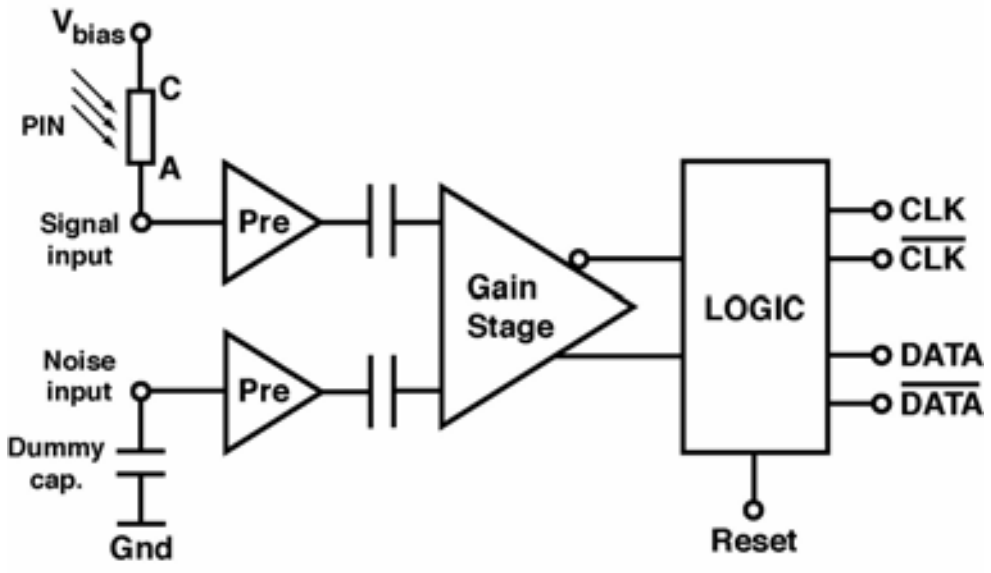

Figure 2: Block diagram of the DORIC circuit.

## 4. Results From IBM 0.25 micron Submissions

The pixel detector design of the VDC and DORIC takes advantage of the development work for similar circuits [2] used by the outer detector, the SemiConductor Tracker (SCT). Both SCT chips attain radiation-tolerance by using bipolar integrated circuits (AMS 0.8 µm BICMOS) and running with high currents in the transistors at 4 V nominal supply voltage. These chips are therefore not applicable for the higher radiation dosage and lower power budget requirements of the pixel detector.

We originally implemented the VDC and DORIC in radiation-hard DMILL 0.8 µm technology with a nominal supply voltage of 3.2 V. An irradiation study of the DMILL circuits in April 2001 with 24 GeV protons at CERN showed severe degradation of circuit performance. We therefore migrated the VDC and DORIC designs to the standard deep submicron (0.25 µm) CMOS technology which had a nominal supply voltage of 2.5 V. Employing enclosed layout transistors and guard rings [3], this technology was expected to be very radiation hard. We had five prototype runs using 3-metal layers over the course of two years with IBM as the foundry. For the engineering run, the layouts were converted to 5-metal layouts in order to share the wafers with the MCCs for cost saving.

We have extensively tested the chips to verify that they satisfy the design specifications. Figure 3 shows the VCSEL current generated by the VDC as a function of the external control current Iset. The saturation at high Iset is due to the large serial resistance of the VCSEL. The dim current is close to the design value of 1 mA. The performance of the chips on opto-boards has been studied in detail. The typical PIN current thresholds for no bit errors are low, ~15-40 µA.

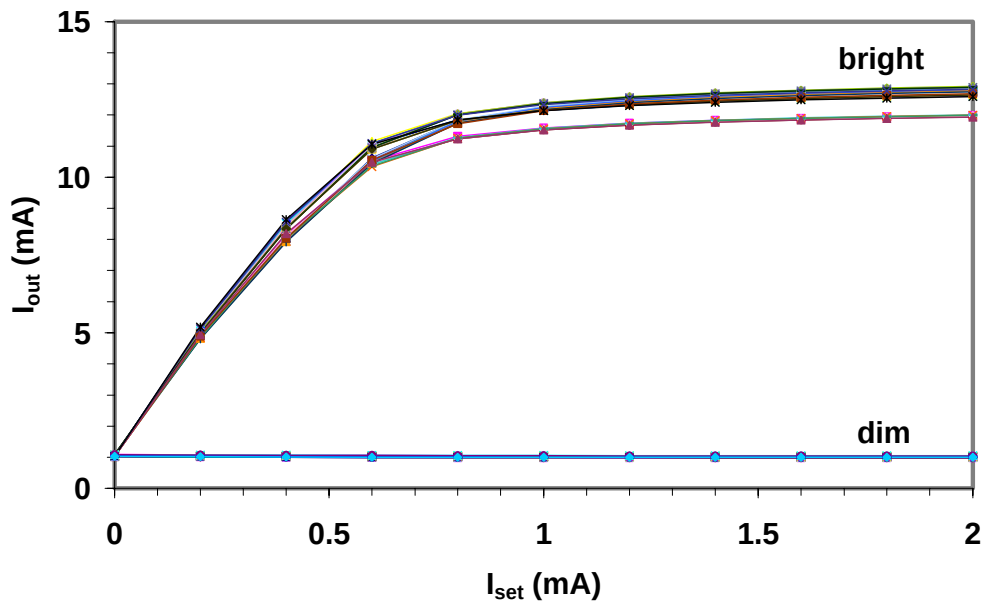

Figure 3: VCSEL drive current vs. $I_{set}$ of eight 4-channel VDCs.

## 5. Irradiation Studies

In the last four years, we have performed four irradiations of the VDCs and DORICs produced using the deep submicron process. We used 24 GeV protons at CERN for the study.

We used two setups in the first three irradiations. In the first setup (“cold box”), we perform electrical testing of VDCs and DORICs. No optical components (VCSEL or PIN) are used, allowing direct study of the possible degradation of the chips without additional complications from the optical components. For the VDCs, we monitor the rise and fall times and the bright and dim currents. For the DORICs, we monitor the minimum input signal for no bit errors, clock jitter and duty cycle, rise and fall times and amplitude and average of the clock and command LVDS.

Four DORICs and four VDCs were irradiated in the cold box setup. We observe no significant degradation in the chip performance up to a total dose of ~62 Mrad, except the average of the clock LVDS of one DORIC increases by 10%, an acceptable change.

In the second setup (“shuttle”), we tested the performance of the opto-link using opto-boards. In the control room, we generated bi-phase mark encoded pseudo-random signals for transmission via 25 m of optical fibers to the opto-boards. The PIN diodes on the opto-boards converted the optical signals into electrical signals. The DORICs then decoded the electrical signals to extract the clock and command LVDS. The LVDS were fed into the VDCs and converted into signals that were appropriate to drive the VCSELs. The optical signals were then sent back to the control room for comparison with the generated signals. We remotely moved the opto-boards on the shuttle out of the beam to anneal the VCSELs. We typically irradiated the opto-boards for 5 hours (~5 Mrad) and then annealed the VCSELs for the rest of the day with large current (~13 mA in the last two irradiations).

We used only the shuttle setup in the last irradiation in order to study the optical link with the latest prototype opto-boards. Four opto-boards were irradiated with the dosage of up to 1.2 x $10^{15}$ p/cm$^2$ (33 Mrad). The PIN current thresholds for no bit errors were all below 40 µA and remained constant through out the irradiation as shown in Fig. 4.

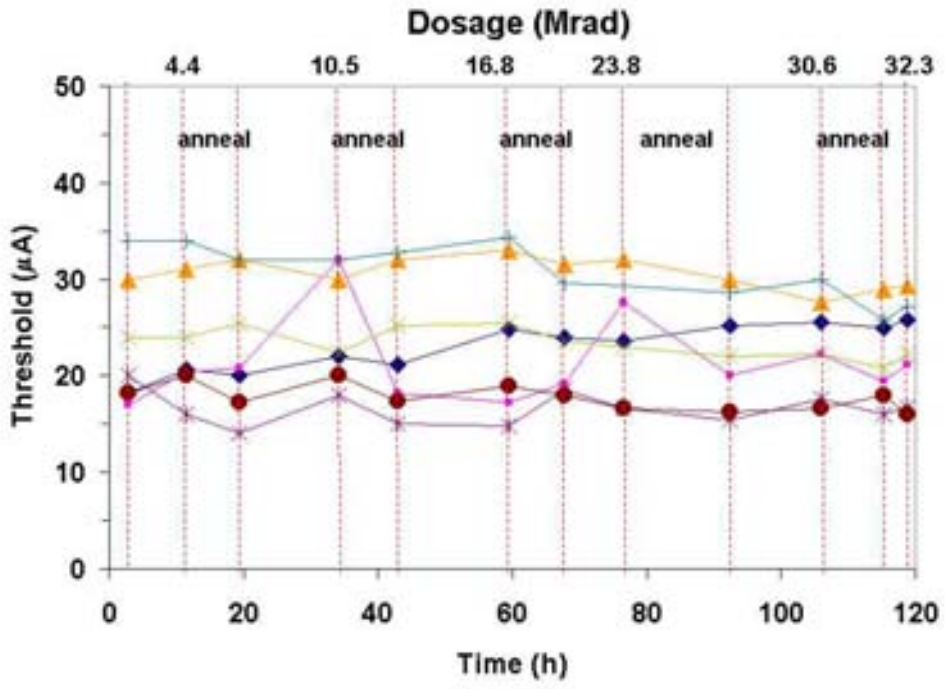

Figure 4: PIN current thresholds for no bit errors as a function of time (dosage) for one of the opto-boards with seven active links in the shuttle setup.

We have converted the observed bit errors due to proton induced current in the PIN diodes into the expected bit error rate at the optical link location in the ATLAS detector as shown in Fig. 5. The bit error rate decreases with increasing PIN current threshold as expected. The bit error rate is ~ 3x$10^{-10}$ at 100 µA, significantly above that achieved with a stand alone DORIC.

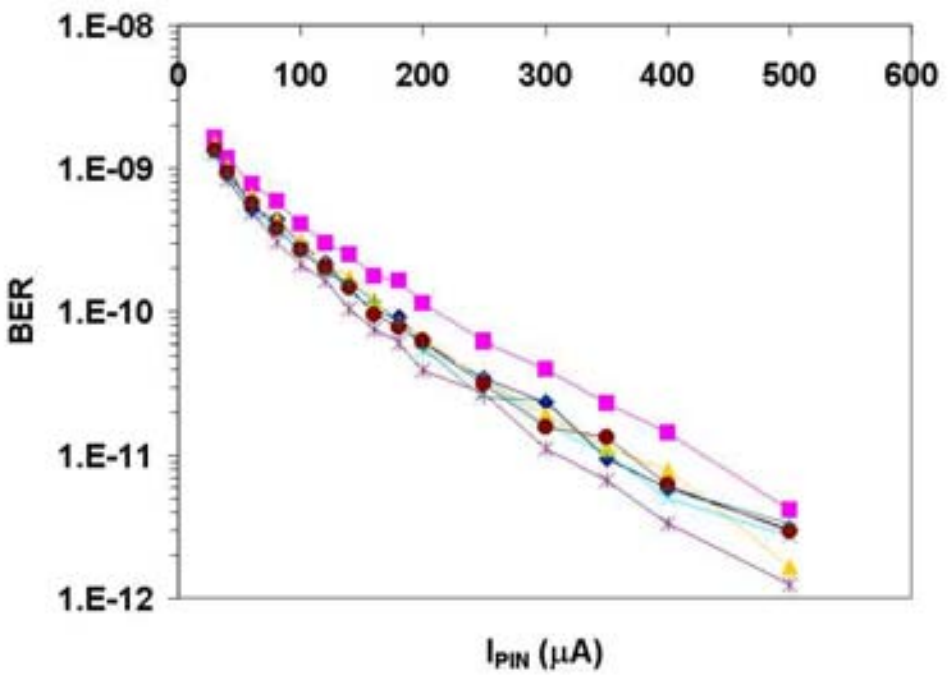

Figure 5: Expected bit error rate as a function of PIN current threshold at the opto-link location in the ATLAS detector.

The optical power from the opto-boards was monitored in the control room during the irradiation. Figure 6 shows the optical power as a function of time (dosage) for one of the opto-boards. We observed a general trend in the data: during the irradiation the optical power decreased; the optical power increased during the annealing, as expected. The power loss was due to radiation damage to the VCSELs as the VDCs and DORICs showed no radiation damage with up to ~62 Mrad of irradiation in the cold-box setup. The optical power of all links was well above 350 µW, the specification for absolute minimum power after irradiation.

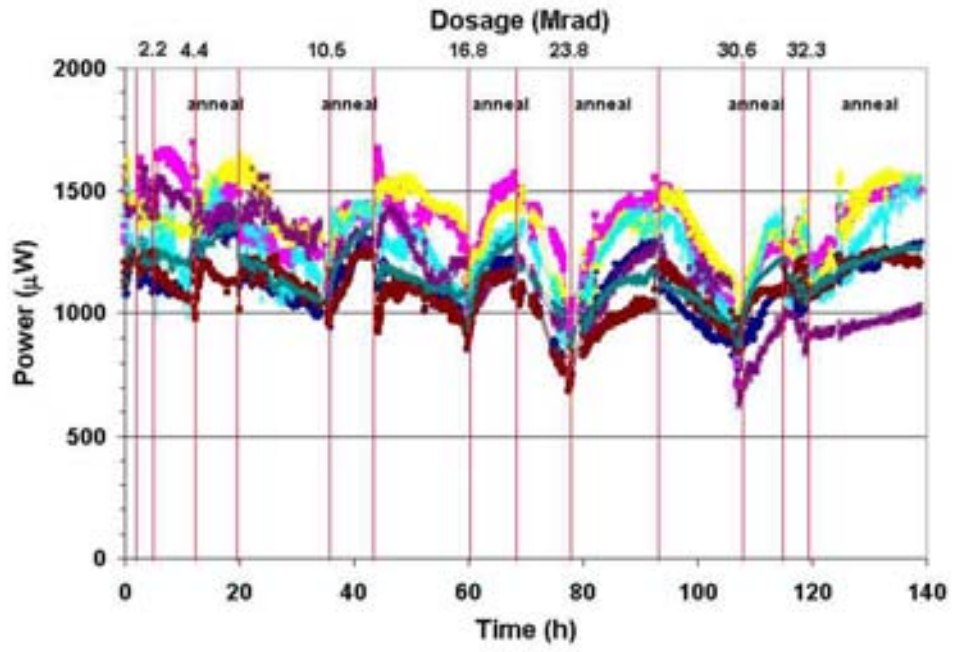

Figure 6: Optical power as a function of time (dosage) for one of the opto-boards with seven active links in the shuttle setup.

## 6. Summary

We have developed the on-detector optical link for the ATLAS pixel detector. The link meets all the requirements and further appears to be sufficiently radiation hard for ten years of operation at the LHC.

## Acknowledgements

This work was supported in part by the U.S. Department of Energy under contract No. DE-FG-02-91ER-40690 and the German Federal Minister for Research and Technology (BMBF) under contract 056Si74.